\documentclass[11pt]{article}
\usepackage{osajnl}
\begin{document}
\title{Spectrally resolved optical frequency comb from a self-referenced 5 GHz femtosecond laser\footnote{contribution of an agency of the US
government, not subject to copyright}}
\author{A. Bartels, R. Gebs}
\address{Gigaoptics GmbH, Blarerstrasse 56, 78462 Konstanz, Germany
and\\ University of Konstanz, Universit\"atsstrasse 10, 78457
Konstanz, Germany} \email{bartels@gigaoptics.com}
\author{M. Kirchner, and S.A. Diddams}
\address{National Institute of Standards and Technology, 325 Broadway M.S. 847, Boulder, CO 80305, USA}
\begin{abstract}
We report a mode-locked Ti:sapphire femtosecond laser with 5\,GHz
repetition rate. Spectral broadening of the 24\,fs pulses in a
microstructured fiber yields an octave-spanning spectrum and
permits self-referencing and active stabilization of the emitted
femtosecond laser frequency comb (FLFC). The individual modes of
the 5\,GHz FLFC are resolved with a high-resolution spectrometer
based on a virtually imaged phased array (VIPA) spectral
disperser. Isolation of single comb elements at a microwatt
average power level is demonstrated. The combination of the
high-power, frequency-stabilized 5\,GHz laser and the
straightforward resolution of its many modes will benefit
applications in direct frequency comb spectroscopy. Additionally,
such a stabilized FLFC should serve as a useful tool for direct
mode-by-mode Fourier synthesis of arbitrary optical waveforms.
\end{abstract}

Basic and applied research in the fields of optical frequency
metrology \cite{udem02}, direct frequency comb spectroscopy
\cite{marian05,gerginov06}, low-noise microwave generation
\cite{bartels05}, and arbitrary waveform generation
\cite{Jiang05b} all benefit from the large mode spacing and high
power per mode available from stabilized femtosecond laser
frequency combs (FLFC) operating at GHz repetition rates. Indeed,
most frequency comb applications benefit from the highest
repetition rate within the available photodetector bandwidth for
which an octave-spanning spectrum appropriate for self-referencing
can be achieved. Optical frequency metrology and direct frequency
comb spectroscopy specifically take advantage of the high power
per mode since in a typical experiment only the power contained in
a few comb teeth contribute to the measured signal, while the
other teeth within the detected optical bandwidth add to the
measurement noise. Thus, a higher repetition rate intrinsically
permits a higher signal-to-noise ratio (S/N). Arbitrary waveform
generation experiments aimed at individually addressing the FLFC
elements in amplitude and phase via spatial light modulators
benefit because a larger mode spacing reduces the required
resolving power for the spectral disperser.

Towards this goal, fundamentally modelocked femtosecond lasers
with repetition rates up to 4\,GHz have been demonstrated \cite
{bartels99,leburn04}. However, the highest reported repetition
rate for a fundamentally modelocked and self-referenced FLFC is
1.4 GHz \cite{fortier06}, with the maximum repetition rate
ultimately limited by the pulse energy required to achieve
sufficient optical bandwidth for self-referencing. Harmonically
modelocked femtosecond and picosecond sources with repetition
rates as high as 10\,GHz and fundamentally modelocked picosecond
lasers with up to 77\,GHz repetition rate have been demonstrated
\cite{tamura01,quinlan06,collings98,zeller07}. However,
self-referencing of such sources could not yet be demonstrated,
thus precluding their application for precision spectroscopy. In
addition, harmonically modelocked sources exhibit residual modes
at the fundamental cavity frequency \cite{quinlan06} posing a
limit to the obtainable modulation contrast in optical arbitrary
waveform generation experiments.

Here, we report a mode-locked femtosecond Ti:sapphire laser
producing 24\,fs pulses at a 5\,GHz repetition rate.
Octave-spanning spectra are produced via spectral broadening in a
small-core microstructured optical fiber, thus enabling the system
to be self-referenced and frequency stabilized. Moreover, the
5\,GHz mode spacing is sufficiently large to enable the individual
modes to be spectrally and spatially separated and recorded in an
efficient two-dimensional format. Direct access to numerous
individual comb modes in a parallel architecture provides unique
capabilities for novel high-resolution spectroscopic techniques
\cite{diddams07} as well as the possibility to precisely control
the amplitude and phase of individual comb modes for the
generation of arbitrary optical and microwave waveforms
\cite{Jiang05b}. The combination of the 5 GHz comb demonstrated
here with line-by-line pulse shaping techniques should ultimately
enable the generation of user-designed optical waveforms that
possess the femtosecond timing jitter available from
well-stabilized optical frequency combs \cite{bartels03}.

The laser cavity is based on a ring design previously used for
repetition rates up to 2\,GHz (see Fig.~\ref{fig1} for schematic)
\cite{bartels99}. The 1.5\,mm long Ti:sapphire crystal is pumped
by 7.5\,W from a 532\,nm laser through a 30\,mm focal length lens
and mirror M1. The focusing mirrors M1 and M2 next to the laser
crystal have a radius of curvature of 15\,mm. The ring cavity is
completed by mirror M3 and the output coupler OC. The cavity
length is 6\,cm, yielding a repetition rate of 5\,GHz. Mirrors
M1-3 have a high-reflective negative dispersive coating with
approximately -40\,fs$^2$ group-delay dispersion (GDD) between
750\,nm and 850\,nm. Together with the laser crystal contribution,
the net cavity GDD is thus approximately -35\,fs$^2$. The output
coupler has 2\% transmission. To initiate mode-locking, mirror M2
is brought close to the inner edge of the cavity stability range
and a slight perturbation is imposed on the cavity (e.g. tapping
on a mirror). When mode-locked, the laser operates
unidirectionally in a random direction and yields 1.15\,W of
output power. We choose the operating direction as indicated in
Fig.~\ref{fig1}. An interferometric autocorrelation trace of the
pulses is shown in Fig.~\ref{fig1}a. The full-width at
half-maximum (FWHM) of the low-pass filtered trace is 36\,fs
corresponding to a pulse duration of 24\,fs under the assumption
of a $\rm{sech^2}$ pulse envelope. The output spectrum is shown in
Fig.~\ref{fig1}b and has a FWHM of 35\,nm centered around 798\,nm.
The 5 GHz repetition rate is detected by focusing $\sim$20\,mW of
optical power onto a high-speed GaAs-pin photodiode. The
electrical power contained in the 5\,GHz signal amounts to -5.7
dBm (into a 50\,$\Omega$ load) with a direct current of
$\sim$5\,mA. This signal is used in a phase-locked loop to
stabilize the repetition rate to an external hydrogen maser
referenced RF-signal at 5.000994 GHz by controlling the cavity
length via a piezo crystal that supports mirror M3.

The carrier-envelope offset frequency $f_0$ of the laser is
measured in an f-2f nonlinear interferometer \cite{jones00}.
950\,mW of the output power is launched into a microstructured
fiber (1.5\,$\mu$m core, zero-GDD wavelength at 590 nm) with an
efficiency of 35\%. The optical output spectrum is shown in
Fig.~\ref{fig1}b. A beat signal between the frequency-doubled
light at 1000\,nm and the fundamental light at 500\,nm is detected
with a Si pin photodiode (see Fig.~\ref{fig2}a). The bandwidth of
the photodiode was approximately 3\,GHz, thus the peaks at $f_R$
and $f_R-f_0$ are partly suppressed. $f_0$ is stabilized at
250\,MHz by controlling the 532\,nm pump power via an
acousto-optic modulator. The stabilized $f_0$ spectrum is shown in
Fig.~\ref{fig2}b. Residual frequency deviations of $f_0$ have been
measured using a high-resolution counter and amount to 6\,mHz at
1\,s gate time.

Approximately 120\,mW of the laser output were split off before
the microstructured fiber and dispersed in a high-resolution
spectrometer that consists of a virtually imaged phased array
(VIPA) spectral disperser orthogonally combined with a
conventional diffraction grating \cite{xiao04,xiao06,diddams07}.
The VIPA etalon has a high-reflective coating on the front face
(except for an uncoated entrance window) and a 96\% reflectance
coating for 800\,nm on the output face. The diffraction grating
has 1800\,lines/mm. The spectrally dispersed elements of the
frequency comb are imaged to a first focal plane where the
spatially separated FLFC components may be individually
manipulated with appropriate devices, e.g. a spatial light
modulator (SLM). The spectrometer output is then imaged to a
second plane and recorded with a CCD camera. Figs.~\ref{fig3}(a)
and (b) show a portion of the optical spectrum in the second focal
plane with nothing placed in the first focal plane. The individual
modes of the 5\,GHz frequency comb are well resolved as dots.
Vertical neighbors within the optical frequency 'brush' are spaced
by one repetition rate, horizontal neighbors are spaced by one
free spectral range (FSR) of the VIPA spectral disperser
($\sim$50\,GHz). The spacing of the dots is $\sim$90\,$\mu$m in
the vertical direction and $\sim$60\,$\mu$m in the horizontal
direction. These values are well-suited for spectroscopic
detection \cite{diddams07} and manipulation of individual FLFC
components using two-dimensional SLMs. Here we perform a simpler
proof-of-principle experiment by isolating a single frequency comb
mode using a 50\,$\mu$m diameter pinhole in the first focal plane.
Fig.~\ref{fig3}(c) shows a view of a single isolated mode at
around 802.5\,nm wavelength from the center region of the
spectrometer output. A power measurement behind the pinhole shows
that the isolated mode contains 2.2\,$\mu$W of average power.

We have isolated a series of different modes by scanning the
pinhole across the dot pattern as indicated in Fig.~\ref{fig3}(b).
The isolated light has been coupled into a single mode fiber and
analyzed with a high resolution (7\,GHz) optical spectrum analyzer
(OSA). The spectra of the isolated dots number 1 and 2 (spaced by
one repetition rate) are shown in the inset of Fig.~\ref{fig4}.
Crosstalk from modes that are spaced by one or more horizontal
spacings (i.e. by multiples of the FSR of the VIPA disperser) is
suppressed by more than 20\,dB. Crosstalk from modes that are
spaced by one or more vertical spacings (i.e. by multiples of
$f_R$) cannot be resolved by the spectrometer. It is expected that
this crosstalk is significantly below 20\,dB because the vertical
spatial mode spacing is $\sim$1.5\,times greater than the
horizontal spacing. The center frequencies of the individual modes
are extracted from the OSA measurement and plotted versus mode
number in Fig.~\ref{fig4}. A linear fit to the data yields a value
of 5.09 GHz per mode in good agreement with the stabilized value
of 5.000994 GHz within the error that is given by the frequency
repeatability of the OSA (2.3\,GHz in 1\,minute). It should be
noted that the specified accuracy of the OSA is only 50\,GHz, i.e.
the absolute values given in Fig.~\ref{fig4} have a common error
of this size that is not indicated in the figure.

A major consideration towards even higher repetition rates is the
nonlinear round-trip phase shift $\Phi_{RT}$ experienced by the
pulses in the Ti:sapphire crystal \cite{brabec92}. The minimum
demonstrated $\Phi_{RT}$ for stable operation of a similar laser
was about 50\,mrad \cite{bartels99}. Here, we estimate
$\Phi_{RT}\approx$200\,mrad. Thus, repetition rates well above
5\,GHz should be possible with the presented design. We succeeded
in shortening the cavity to yield 6\,GHz repetition rate with no
significant changes to the output characteristics (see
Fig.~\ref{fig2}c for an RF spectrum of the repetition rate
signal). To our knowledge, this is the highest repetition rate
ever demonstrated for a fundamentally modelocked femtosecond
laser. Mechanical constraints prevented higher repetition rates.
We anticipate that it is straightforward to scale the presented
resonator to at least 10\,GHz.

In conclusion we have demonstrated a femtosecond laser with 5\,GHz
repetition rate that can be stabilized in both repetition rate and
carrier-envelope offset frequency. We are able to spectrally
resolve the emitted FLFC with a VIPA spectral disperser and to
isolate individual elements at a microwatt power level. Optical
frequency metrology, direct frequency comb spectroscopy and
line-by-line pulse shaping will benefit from these experiments.

We thank Andrew Weiner and Leo Hollberg for their vital
contributions to this work and acknowledge thoughtful comments on
this manuscript provided by Tara Fortier and Jason Stalnaker.


\newpage
\begin{figure}[hp]
\caption{(a) Interferometric autocorrelation trace (solid line) of
the 5\,GHz pulse train and the low-pass filtered trace (dashed
line). Inset: Cavity schematic of the laser. (b) Laser output
spectrum scaled in units of power per 5\,GHz mode (dashed line)
and octave-spanning spectrum after broadening in microstructured
optical fiber (solid line). } \label{fig1}
\end{figure}
\newpage
\begin{figure}[hp]
\caption{(a) Self-referencing beat signal of the 5\,GHz laser
showing the carrier-envelope offset frequency $f_0$ and the
repetition rate $f_R$. The resolution bandwidth (RBW) is 300 kHz.
(b) Phase-locked $f_0$ signal with RBW set to 100\,Hz. (c)
Photodetected 6\,GHz repetition rate signal (RBW is 30\,kHz).}
\label{fig2}
\end{figure}
\newpage
\begin{figure}[hp]
\caption{(a) Output of the VIPA spectral disperser recorded with a
CCD camera. The image covers $\sim$5\,nm of the spectrum centered
around 802.5\,nm. (b) Zoom into the CCD image. Successive modes of
the FLFC are numbered. (c) Mode number 1 isolated with a pinhole.}
\label{fig3}
\end{figure}
\newpage
\begin{figure}[hp]
\caption{Frequency of the isolated FLFC modes versus mode number.
Error bars represent the frequency repeatability of the OSA. The
line represents the expected dependence for the 5.000994\,GHz
repetition rate. Inset: Spectra of modes number 1 (solid line) and
2 (dashed line).} \label{fig4}
\end{figure}
\newpage
\centerline{\scalebox{.99}{\includegraphics{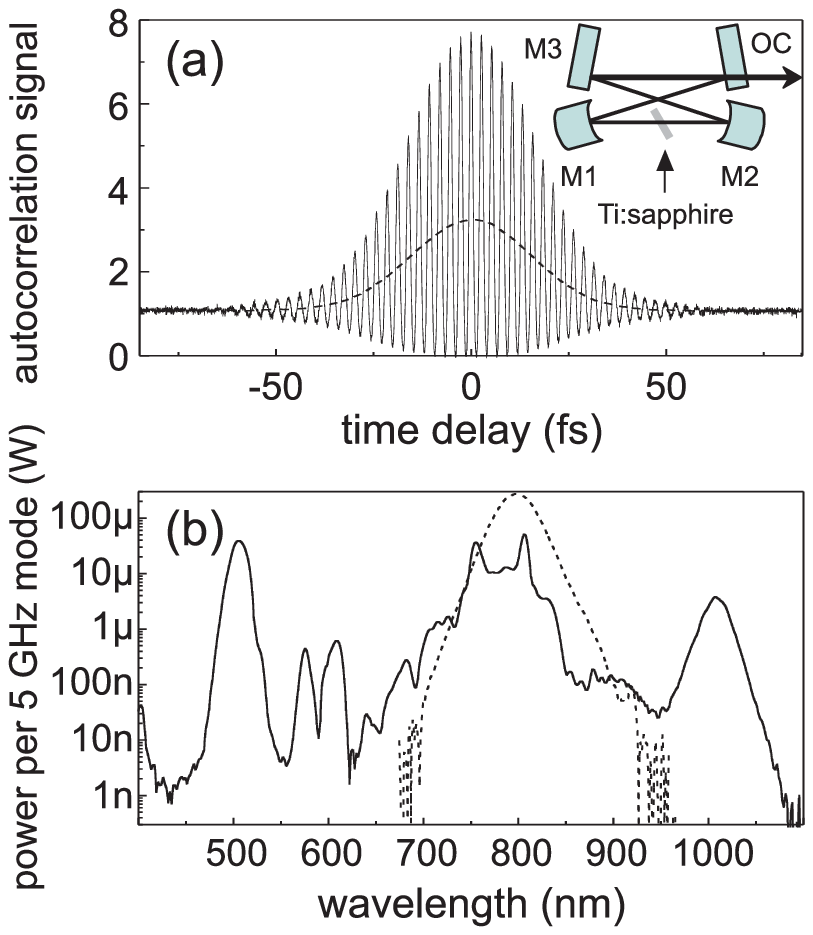}}} \vskip2cm
Figure 1, A. Bartels et al.
\newpage
\centerline{\scalebox{.99}{\includegraphics{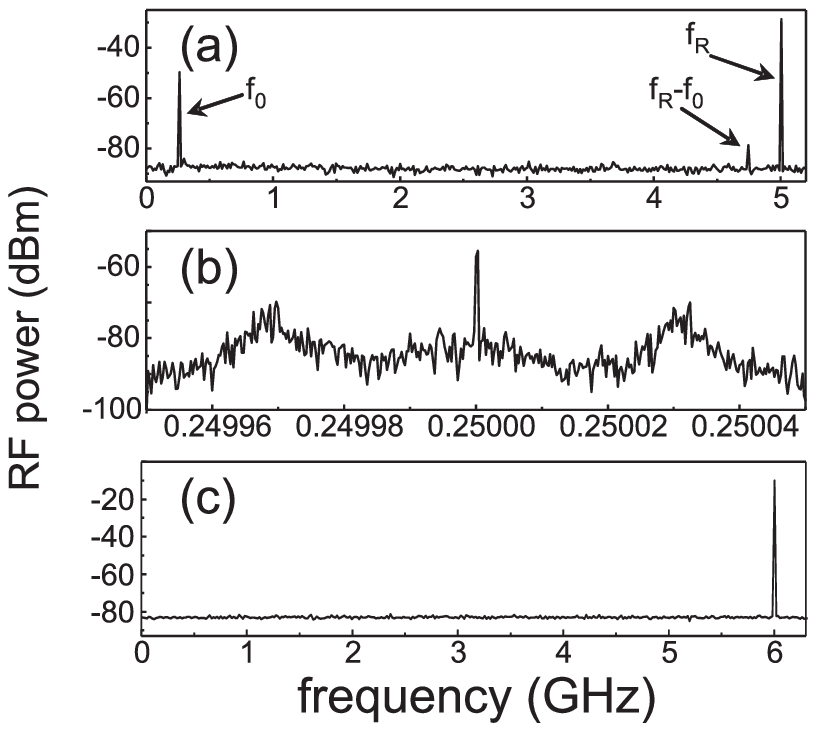}}} \vskip2cm
Figure 2, A. Bartels et al.
\newpage
\centerline{\scalebox{.6}{\includegraphics{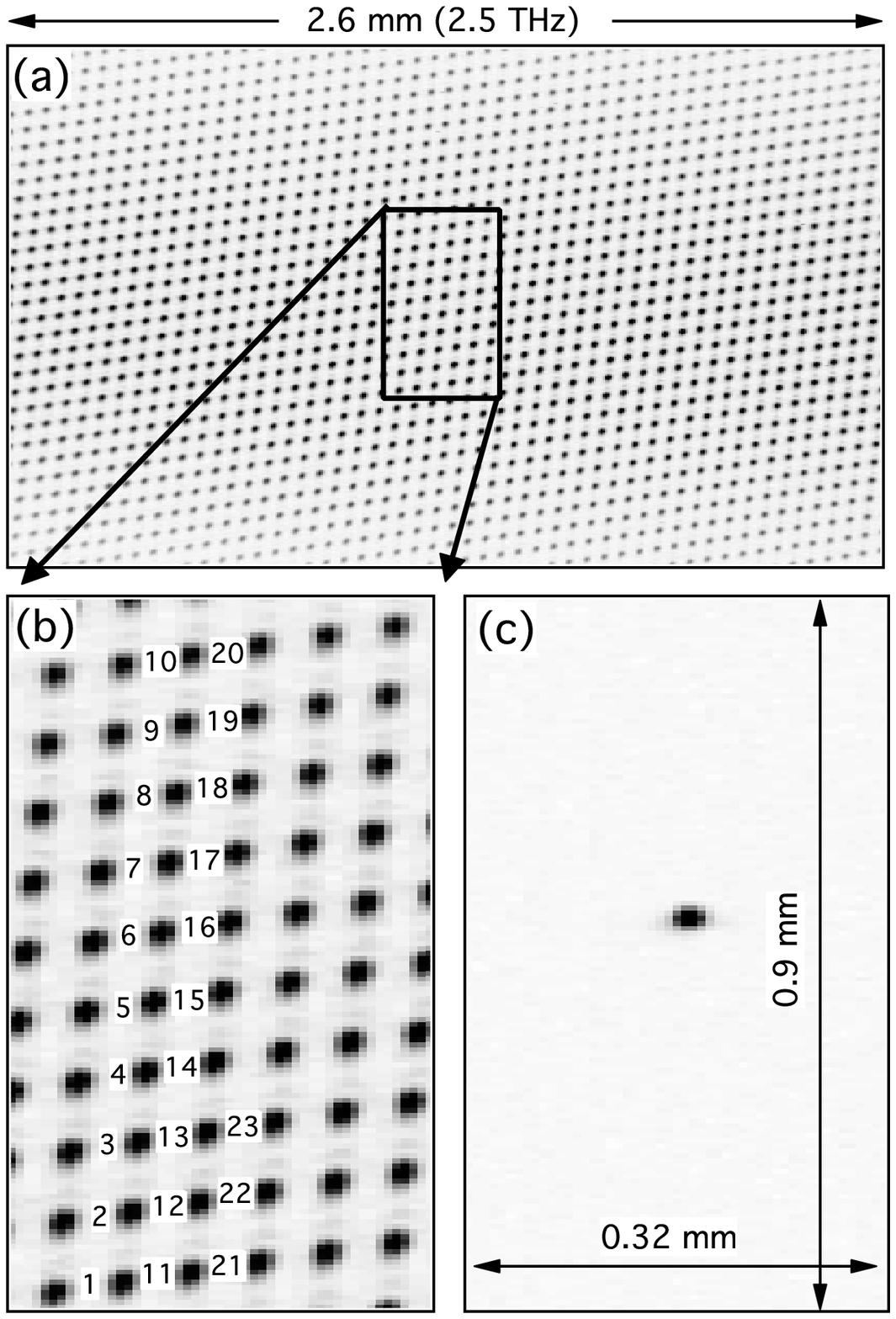}}} \vskip2cm
Figure 3, A. Bartels et al.
\newpage
\centerline{\scalebox{.99}{\includegraphics{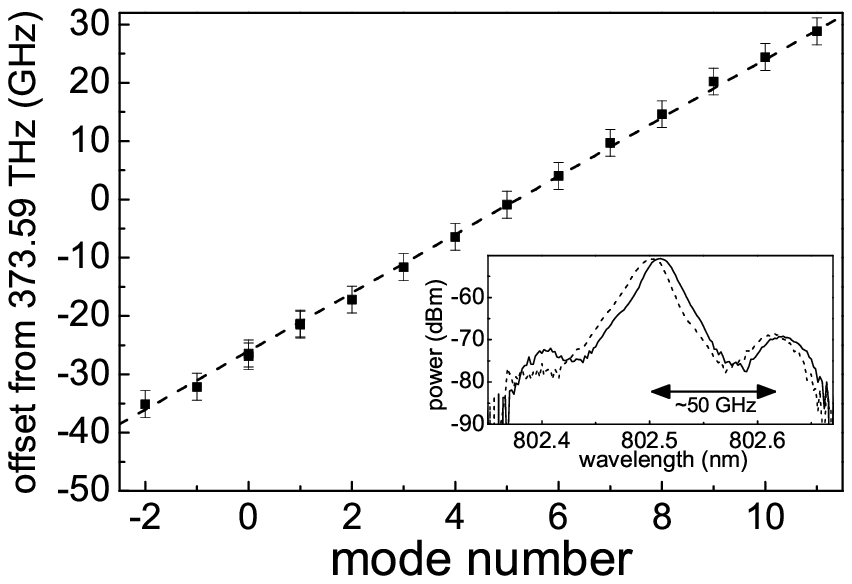}}} \vskip2cm
Figure 4, A. Bartels et al.
\end{document}